\documentclass[a4paper,fleqn,usenatbib]{mnras}

\usepackage[T1]{fontenc}
\usepackage{ae,aecompl}

\usepackage{graphicx}	
\usepackage{amsmath}	
\usepackage{amssymb}	
\usepackage{amsfonts}
\usepackage{url}

\title[Eclipsing binary black hole candidate in Mrk~421]{An eclipsing
binary black hole candidate system in the blazar Mrk~421}

\author[E. Ben\'{\i}tez et al.]{
E.~Ben\'{\i}tez,$^{1}$
J.~I. Cabrera,$^{2,3}$
N.~Fraija,$^{1}$
X. Hern\'andez,$^{1}$
O.~L\'opez-Corona,$^{1,4,5}$
\newauthor
F.~D. Lora-Clavijo,$^{1,6}$
S.~Mendoza.$^{1}$\thanks{Email: sergio@astro.unam.mx}
\\
$^{1}$Instituto de Astronom\'{\i}a, Universidad Nacional
            Aut\'onoma de M\'exico, AP 70-264, Ciudad de M\'exico 04510,
	    M\'exico\\
$^{2}$Escuela Superior de F\'{\i}sica y Matem\'aticas, Instituto
            Polit\'ecnico Nacional, U.P. Adolfo L\'opez Mateos, Ciudad de
	    M\'exico 07730, M\'exico\\
$^{3}$Facultad de Ciencias, Universidad Nacional Aut\'onoma de
            M\'exico, Ciudad de M\'exico 04510, M\'exico\\
$^{4}$Centro de Ciencias de la Complejidad, Universidad Nacional
            Aut\'onoma de M\'exico, Ciudad de M\'exico 04510, M\'exico\\
$^{5}$Instituto de Investigaci\'on sobre Desarrollo Sustentable y Equidad
      Social, Universidad Iberoamericana, Ciudad de M\'exico, México\\
$^{6}$Grupo de Investigaci\'on en Relatividad y Gravitaci\'on,
            Escuela de F\'isica, Universidad Industrial de Santander,
            A. A. 678, \\
	    Bucaramanga 680002, Colombia
}

\date{Accepted XXX. Received YYY; in original form ZZZ}

\pubyear{2015}

\begin{document}
\label{firstpage}
\pagerange{\pageref{firstpage}--\pageref{lastpage}}
\maketitle

\begin{abstract}
  Removing outbursts from multiwavelength light curves of the blazar
Mrk~421, we construct outburstless time series for this system.
A model-independent power spectrum light curve analysis in the optical,
hard X-ray and \( \gamma \)-rays for this outburstless state and also
the full light-curves, show clear evidence for a periodicity of \(
\approx 310~\text{days} \) across all wavelengths studied.  A subsequent
full maximum likelihood analysis fitting an eclipse model confirms
a periodicity of \( 310 \pm 1~\text{days}\).  The power spectrum of
the signal in the outburstless state of the source does not follow
a flicker noise behaviour and so, the system producing it is not
self-organised. This and the fact that the periodicity is better defined
in the outburstless state, strongly suggests that it is not produced
by any internal physical processes associated to the central engine.
The simplest physical mechanism to which this periodicity could be
ascribed is a dynamical effect produced by an orbiting supermassive
black hole companion  eclipsing the central engine.  Interestingly,
the optimal eclipse model infers a brightness enhancement of
\( \left( 136.4 \pm 20  \right) \% \), suggesting an eclipse
resulting in a gravitational lensing brightening.  Consisting with this
interpretation, the eclipse occurs for only \( \left( 9.7 \pm 0.2 \right)
\% \) of the orbital period.
\end{abstract}




\section{Introduction}

Among active galactic nuclei, blazars are considered the most variable sources
since they exhibit the most rapid and largest amplitude continuum
variations extending from radio to X-rays, and many of them up to the
$\gamma$-rays. The detection of variability in blazars is considered a
valuable tool since it enables us to constrain properties
associated with their central emission regions and set upper limits on the
physical scale of the system through causality arguments~\citep[see e.g.][]{1997ARA&A..35..445U}.

 Several variability studies in blazars have been done with the
aim of finding periodic or quasi-periodic variations, which could
establish the presence of supermassive binary black hole
systems~\citep[cf.][]{1980Natur.287..307B}. To date, the best known
example of a binary black hole candidate associated with a blazar is found
in OJ~287, an object which shows a variability with period \( \sim 12
\)~years. The periodicity was discovered by analysing its historical
light curve comprising data for more than 100 years in the optical
V~band \citep[see][]{1988ApJ...325..628S, 1996ApJ...460..207L}.
This result led to the proposal of a precessing binary black hole model for
OJ~287, which can reproduce the periodic outbursts displayed by this
object.  Recently, more blazars have been considered as supermassive black
hole candidates~\citep[cf.][]{Riger2007}, and new results on quasi-periodical
oscillations have been reported by e.g.~\citet{2016AJ....151...54S} and~
\citet{2015ApJ...813L..41A}.

  Direct imaging of binary supermassive black hole systems in active galactic nuclei have
been performed with great success in very few cases~\citep[see e.g.][and
references therein]{2014Natur.511...57D}, ranging from the kiloparsec
to the extreme \( \sim 5 \, \text{pc} \) separation scales.  The small
outburst to quiescent active galactic nuclei fraction raises the possibility of a large
population of binary black holes remaining undetected, and highlights
the importance of developing search techniques for such objects in the
absence of outbursts.

  Mrk~421 is a BL Lac object hosted by an elliptical galaxy  and
one of the most well studied blazars due to its proximity, with  a
redshift \( z \approx 0.03 \) \citep{1975ApJ...198..261U}.  
Although several multiwavelength variability
studies have been done on Mrk~421\citep[see][and references
therein]{2011ApJ...736..131A,2014A&A...570A..77P}, the light-curves obtained  for this
blazar~\citep[e.g.][]{1997A&AS..123..569L,chen14} have not shown fully
convincing evidence of periodic or quasi-periodic variations, potentially
due to shortcomings in the statistical methods employed~\citep{baluev}.

  In this letter we model multiwavelength light curves of Mrk~421 using
the free software R-package RobPer~\citep{anita02}, to analyse the
associated periodograms on time scales of several years.  The method
is applied to what we define as ``\emph{outburstless}'' light-curves,
i.e. those curves where the signal coming from flares or outbursts
have been removed.  More precisely, we analyse the
signal below the 3-\(\sigma\) noise level.  In section~\ref{statistical}
we present the multiwavelength database used for the analysis, and the
corresponding periodograms obtained using RobPer.   Once the existence
of a \( \sim 310~\text{days} \) period was found through the periodogram
analysis, a full maximum likelihood model for a generalised eclipse is
introduced to recover the optimal frequency, phase, eclipse fraction and
duration of the eclipse for the X-ray data.  We interpret the result of a negative
eclipse fraction and its small duration as evidence
for a supermassive black hole companion orbiting
about the central engine, which results in a brightening of the light curve.
Finally, we present a discussion of our results in Section~\ref{discussion}.

\section{Data analysis}
\label{statistical}

  The outburstless multifrequency light curves of the Blazar Mrk~421 were
constructed using average fluxes to find the \(3\)-\(\sigma\) level in
three different wavebands.  The optical long-term light curves  analysed in
this work were built using the database of the {\it American Association
of Variable Star Observers (AAVSO)}, from 1981 April 11 to 2014 July 21
($\sim$ 33 years). Average V-band magnitudes, with no errors reported in
the database, were obtained and converted to flux units in Jy.   
The X ray data from \( 15 \)-\( 50 \)~keV were obtained
using the database of the \textit{ SWIFT-BAT} hard X ray transient
monitor~\citep{Krimm} from 2004 December 22 to 2014 May 3 ($\sim$
10 years).  \textit{Fermi-LAT} $\gamma$-ray fluxes were obtained in the
range \( 0.2 \)--\(300\)~GeV using the public database.  The data used
covers the interval 2008 August 08 up to 2014 May 31 ($\sim$ 6 years),
and were reduced following the procedure described in~\citet{nacho}.

  Once the data above the \( 3 \)--\( \sigma \) level are identified, it is
replaced by blank intervals in the light curve.  This generates the
outburstless curves we further analyse.  Thus, we study the signal which
is usually discarded in studies of active galactic nuclei variability.

  An important task in several astrophysical studies is the detection of
periodicities in irregularly sampled time series, or data with low signal
to noise ratios.  Any  periodic behaviour
present in the observed data will be the result of any intrinsic
variations of the system analysed, convolved with any periodicity
imposed by the temporal sampling of the various data observed.  The use
of three distinct energy bands, one from ground based observations and
two from independent satellites, allows a handle on this aspect, as the
time sampling of each is also independent and distinct. The physical
relevance of any feature found in the frequency analysis will depend on
whether it appears in all bands or not, and on whether it coincides or not
with elements imposed by the respective spectral window functions of the 
data used \citep[cf.][]{dawson10}.

  The main problem associated with these measurements is that the classic
Fourier periodogram analysis cannot be applied to irregularly sampled
databases~\citep[e.g.][]{Thieler2013b}.  For the same reasons, the
\citet{Deeming1975fourier} periodogram is not adequate as it is
known to react to periodicities in the sampling \citep[see
e.g.][]{hall2000nonparametric}.
Hence, periodicity searches in this
kind of time series have become a very active research field. In the case
of light curves, the epoch
folding periodogram \citep{leahy1983searches} or the analysis of variance
periodogram \citep{schwarzenberg1989advantage} can be interpreted as fitting
a step function to a light curve. The other choice of preference is to use
sine function implementations e.g. Lomb-Scargle~\citep{Scargle1982} or the 
phase dispersion minimisation periodogram~\citep{Stellingwerf1978}.   In
principle, both classes work equivalently for unevenly sampled time series.
More recent methods use uncentred data and fit a model with
intercept \citep[e.g.][]{cumming1999lick,zechmeister2009generalised}.
The R package RobPer~\citep{Thieler2013b},  incorporates more complex periodic functions in the fitting process such
as periodic splines~\citep[e.g.][]{1996Natur.383..319G,oh2004period}
which provide more robust results.
RobPer applies an outlier search on the periodogram,
instead of using fixed critical values that are theoretically only
justified in case of least squares regression. Among other special
features, RobPer has a very complete pool of regression techniques,
consisting of the classic least squares, least absolute deviations,
least trimmed squares, M-, S- and $\tau$ -regressions.  It 
optionally takes observational uncertainties into
account using weighted regressions.  
As with any other type of periodogram the output represents
the coefficient of determination for each trial frequency.

  Due to the advanced capabilities of the RobPer package and the power
and generality of the R project for Statistical Computing, we have
analysed the multiwavelength light curves of Mrk~421 for the first time
using this software.  For this particular case, we use splines as test
functions and the M-regression using the Huber function.

 The top row of Figure~1 shows the full light curves analysed in
the  optical, X and \(\gamma\)-ray bands. The average flux level is
shown by the solid horizontal line, and the 3-$\sigma$ level by the
dashed horizontal one, which defines the cut taken to construct the
outburstless states.  The middle row of the Figure presents the spectral window
functions associated with the time structure of the sampling in the
respective data sets. The first panel, corresponding to the optical data,
is clearly dominated by a strong feature at a period of 365 days, inherent
to the daytime appearance of Mrk~421 for approximately four months a year.
The first significant peak in the X-ray spectral window function, at a period of
120 days, is produced by the three month gap common to many {\it SWIFT}
sources \citep[e.g.][]{swift-x-ray}. A series of peaks follow at well
defined periods, including again a strong yearly feature. In the $\gamma$
spectral window function we see a sequence of well defined peaks, crucially not
coinciding with the ones in the X-rays. It is clear from the plots in
the middle row of Figure~1 that the time structure of the sampling in
the three bands used is highly independent.

  The bottom row of Figure~1 shows the periodograms for the optical, X
and $\gamma$-ray bands, respectively. In the optical, a number of peaks
are evident, the largest at a period close to 310 days, distinct from
the subsequent feature at 365 days. This last is a reflection of the
time sampling structure evident in the middle row.  A series of narrow
peaks with periods below 190 days also appear in the \( \approx 30 \)
years covered by this data sample.  The middle panel, a periodogram
for the \( \approx 10 \) years covered by the {\it  SWIFT} data, has
as its most prominent element a well defined signal at \( \approx 310
\)~days. Again, this is separate from the small features associated to
the yearly enhancement in the corresponding window function. Other than
this, a lower peak appears below 250 days. The last plot on the bottom
row of the Figure shows the much shorter temporal extent of the \(
\approx 6 \)~years in the $\gamma$-ray  data yielding a prominent and
broadened feature closely centred on 310 days, and two narrow peaks at
160 and 130 days.

Comparing particularly the left and middle plot of the bottom row in 
Figure~1
to their respective window functions, it is clear that the features at
310 days do not correspond to peaks imposed by the very distinct time
sampling of the two data sets, as is also the case in the less well
defined periodogram of the shorter time extent $\gamma$-ray data. Thus,
the 310 day period is constant across bands, despite the widely changing
and completely independent time sampling structures and observational
setups. Outside of the period ranges shown in the middle and bottom rows 
of Figure~1 other peaks appear, but much broader, noisier, and not
at fixed positions across observational bands.

  We note that the window functions and periodograms of the full uncut
light curves, including outbursts, differ little from the ones presented,
which show a somewhat more clearly defined time structure. This shows that
the periodicity detected is not associated to the outburst component,
that which is usually analysed, but to the underlying outburstless state.

  In general terms, a periodogram analysis not always yields
straightforward confidence intervals.  This is due to the fact that
there is not a well established technique that considers the non-linear
transformation of uncertainties when converting from the time to the
frequency domains. This is also true when a periodicity is calculated
directly in the time domain for data with low signal to noise ratio.

  The power spectrum shows that the multiwavelength signal does not
follow a flicker \( 1 / f \) noise behaviour.  Instead, a more correlated
brown-like noise signal is detected; meaning that the physical processes
responsible for this periodicity are not related to the central engine,
which is typically associated with a self-organised \( 1 / f \) signal
\citep{per,sole,qsr1,qsr2,boyer2009self,corona2014complex}.  The most
parsimonious interpretation of our results is provided by considering them
as evidence suggesting an eclipsing supermassive binary black hole system.

\begin{figure*}
  \begin{center}
  \includegraphics[scale=0.48]{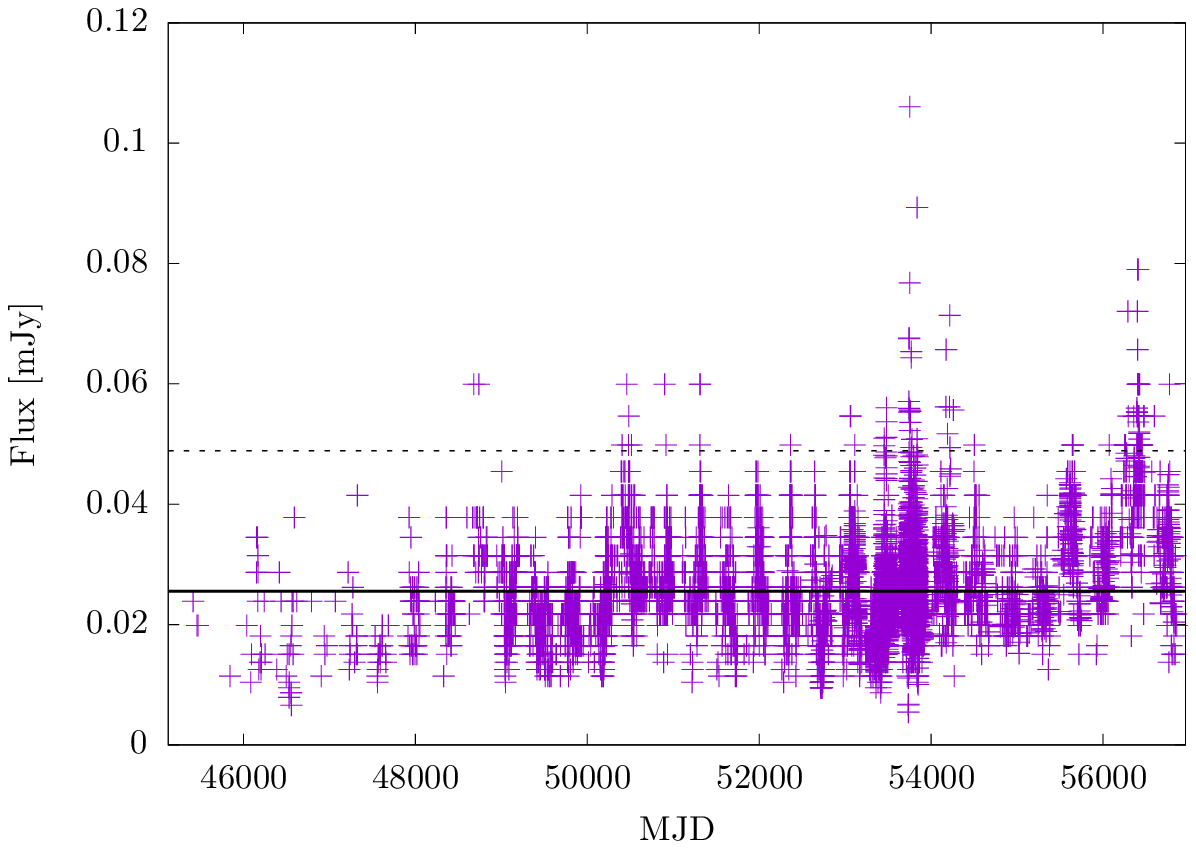}
  \includegraphics[scale=0.48]{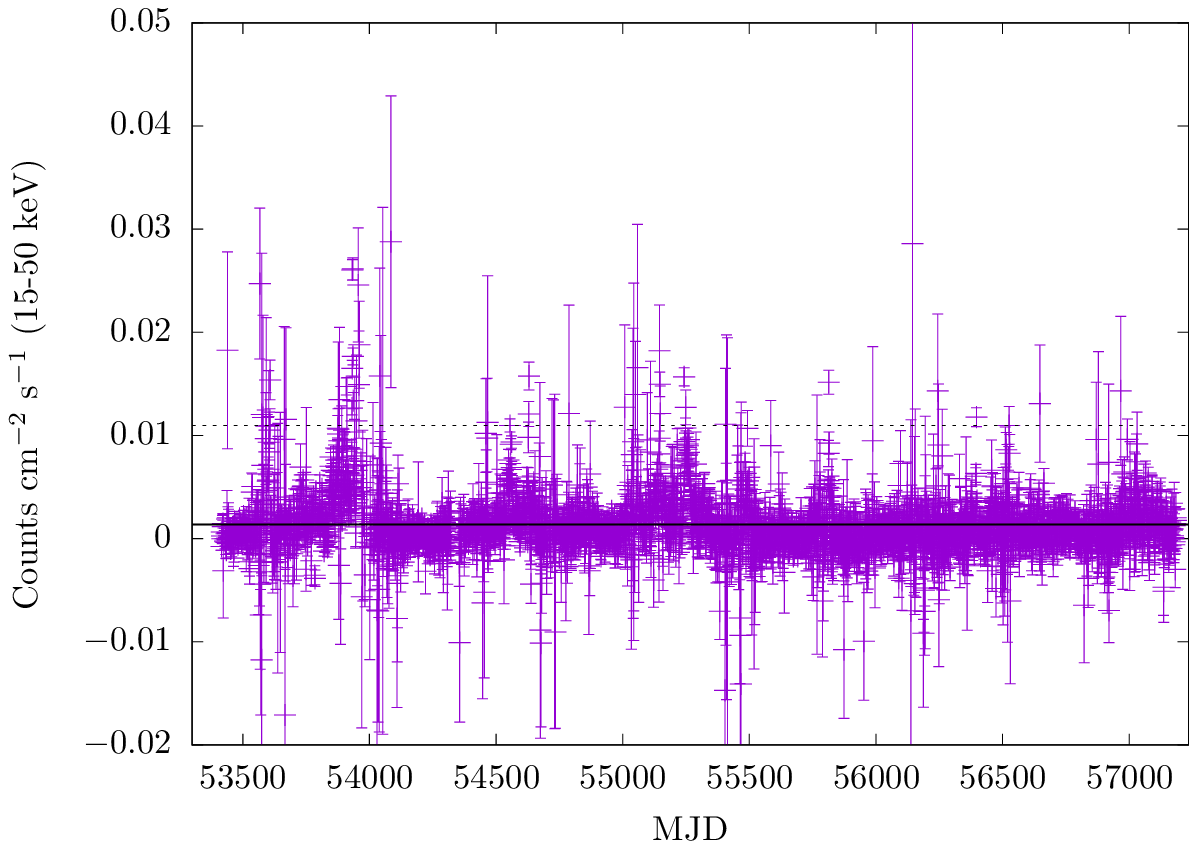}
  \includegraphics[scale=0.48]{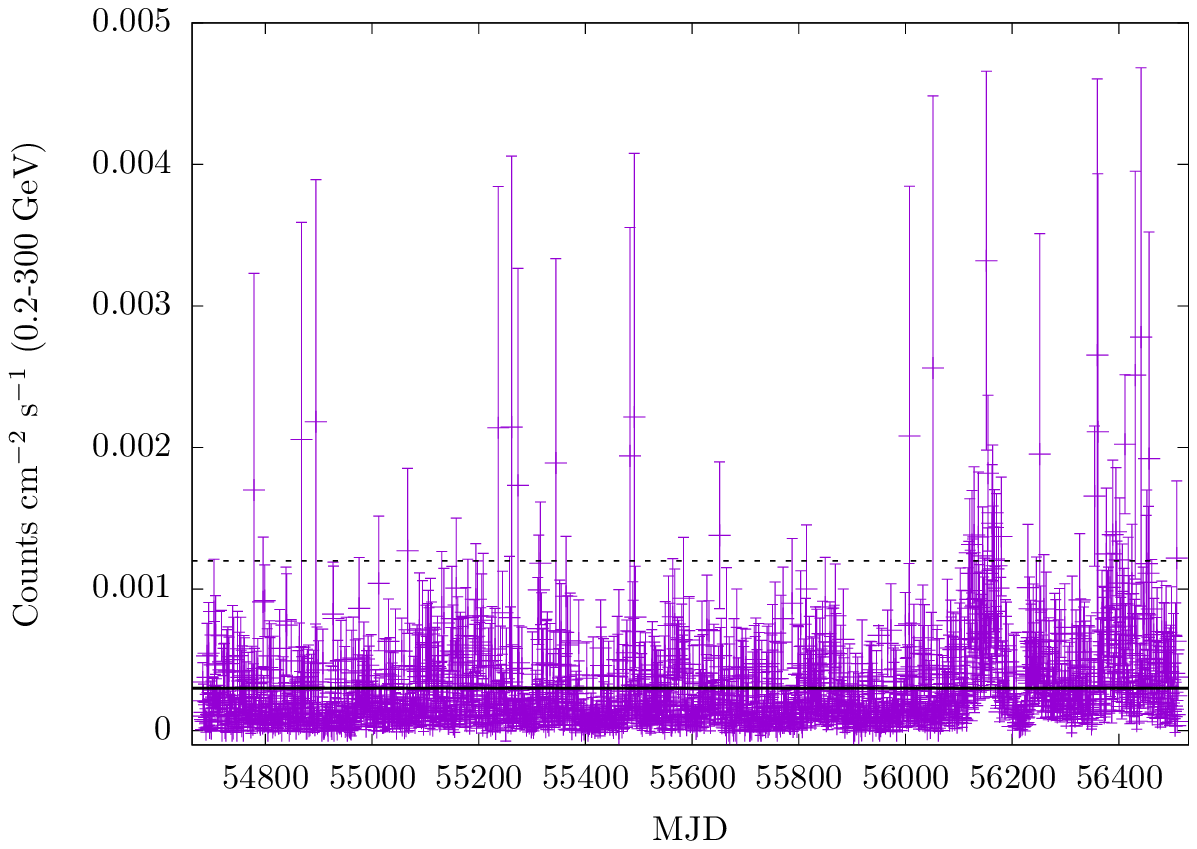} \\
  \includegraphics[scale=0.47]{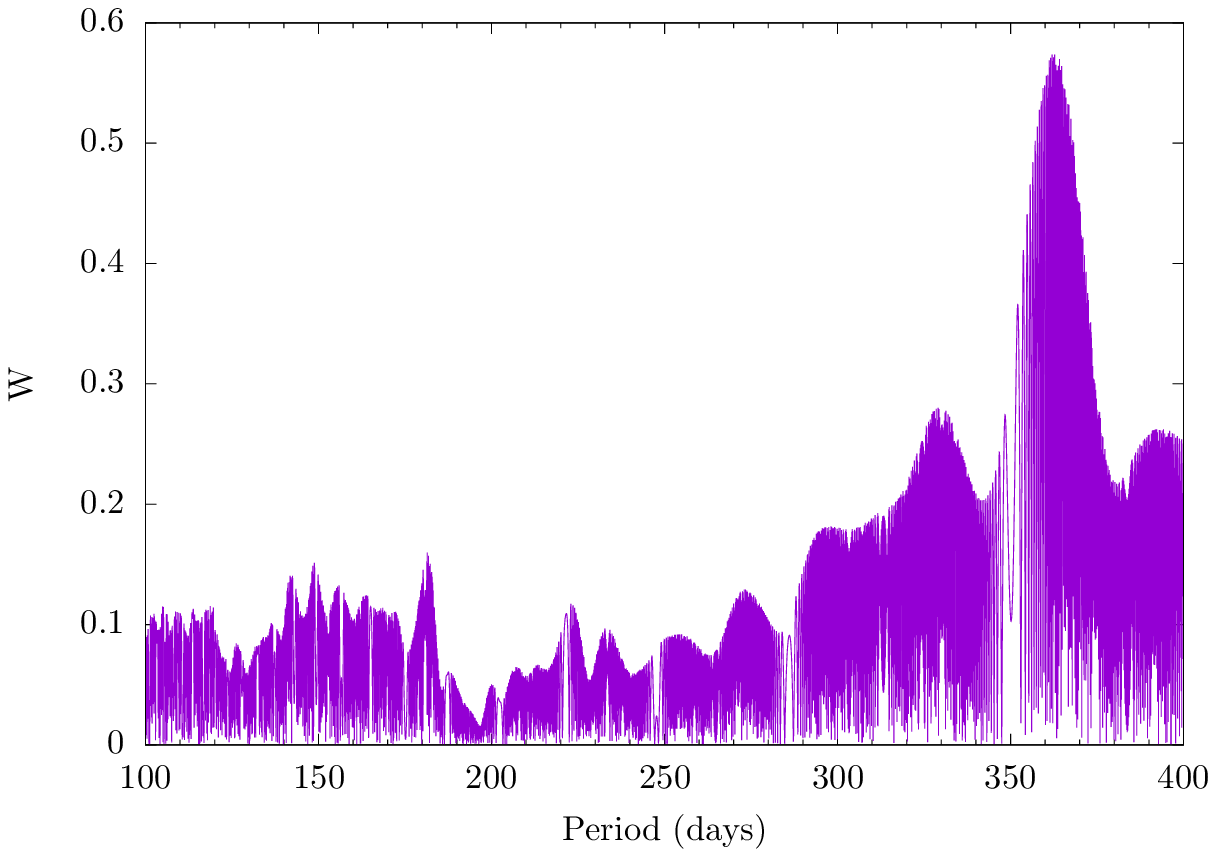}
  \includegraphics[scale=0.47]{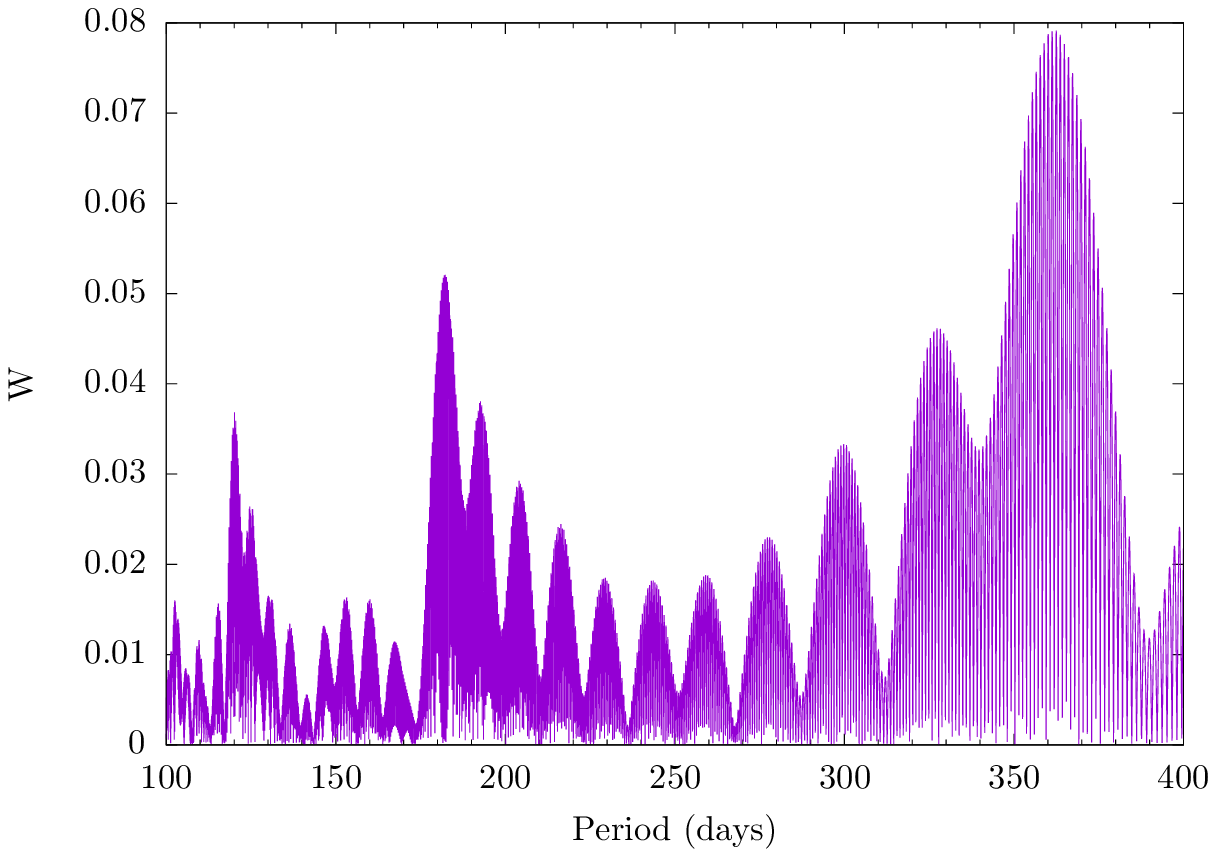}
  \includegraphics[scale=0.47]{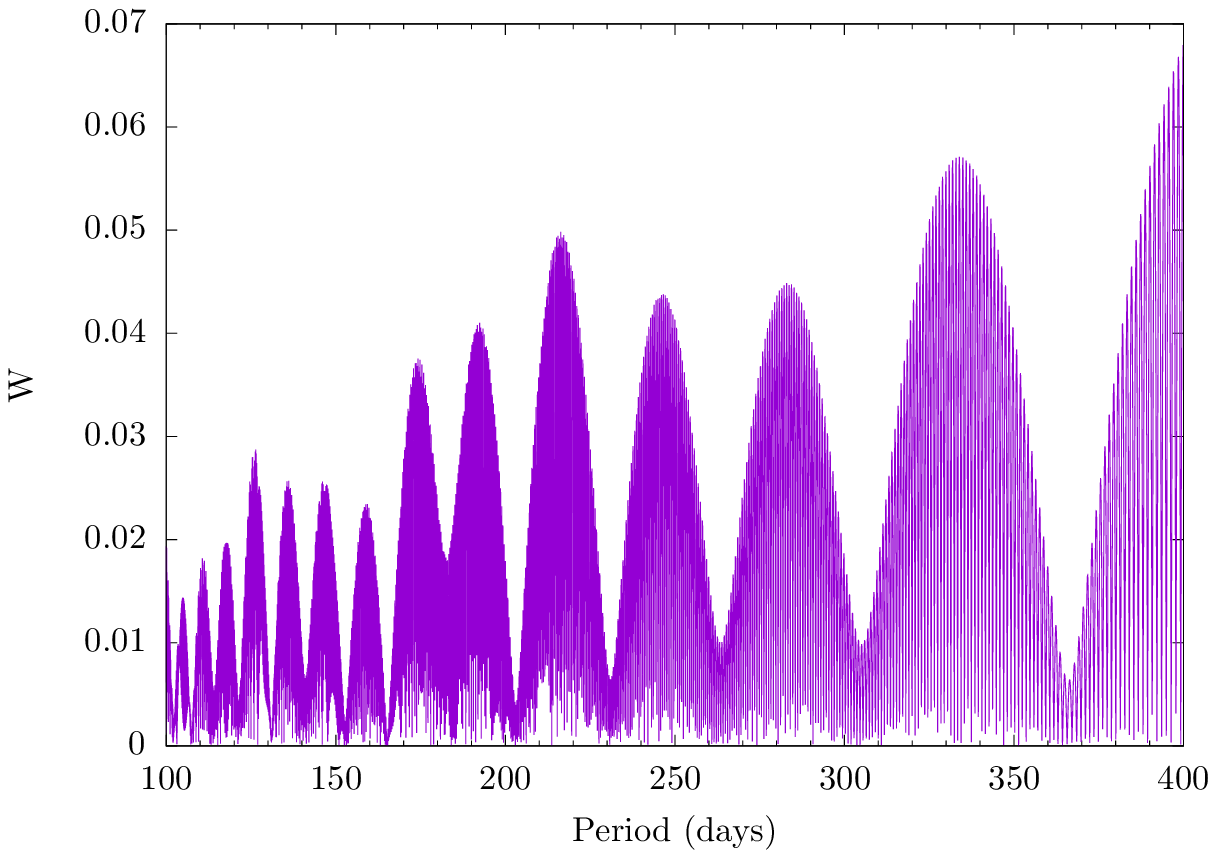} \\
  \includegraphics[scale=0.47]{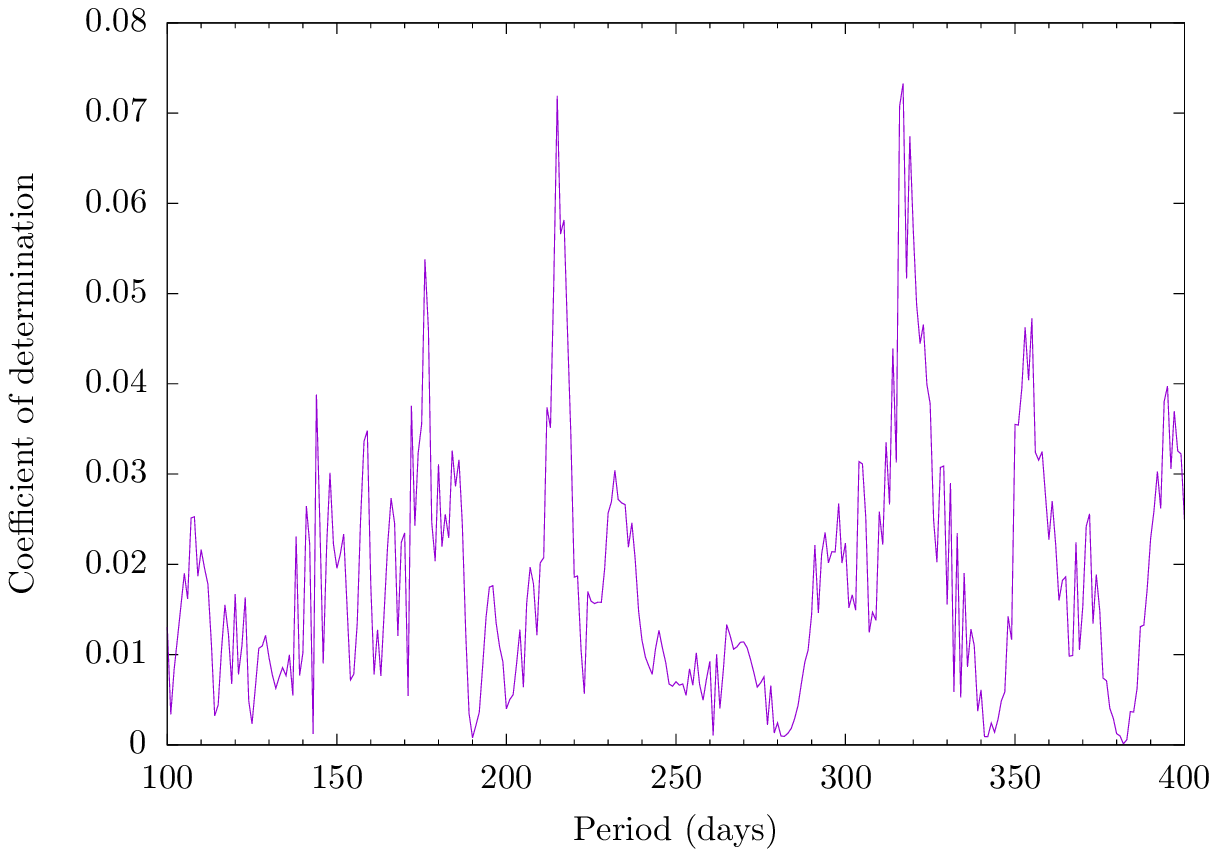}
  \includegraphics[scale=0.47]{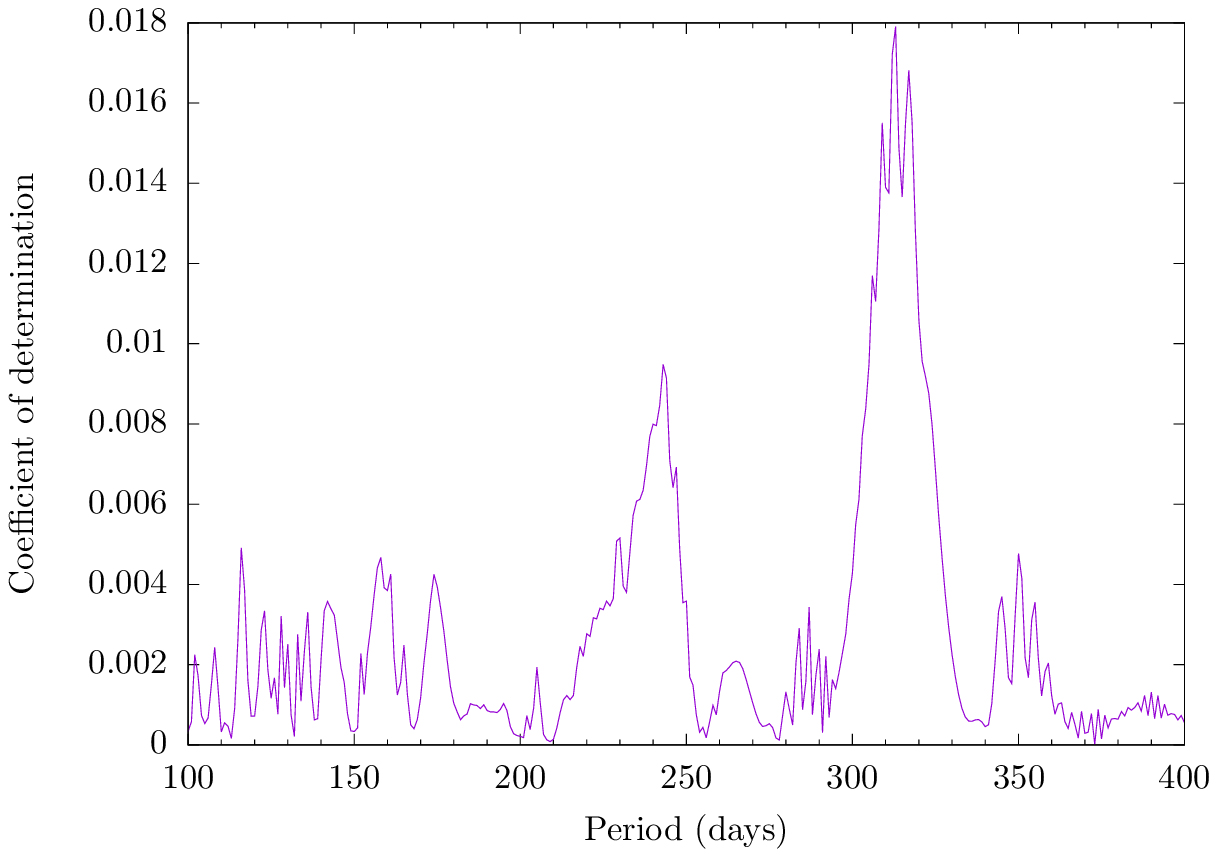}
  \includegraphics[scale=0.47]{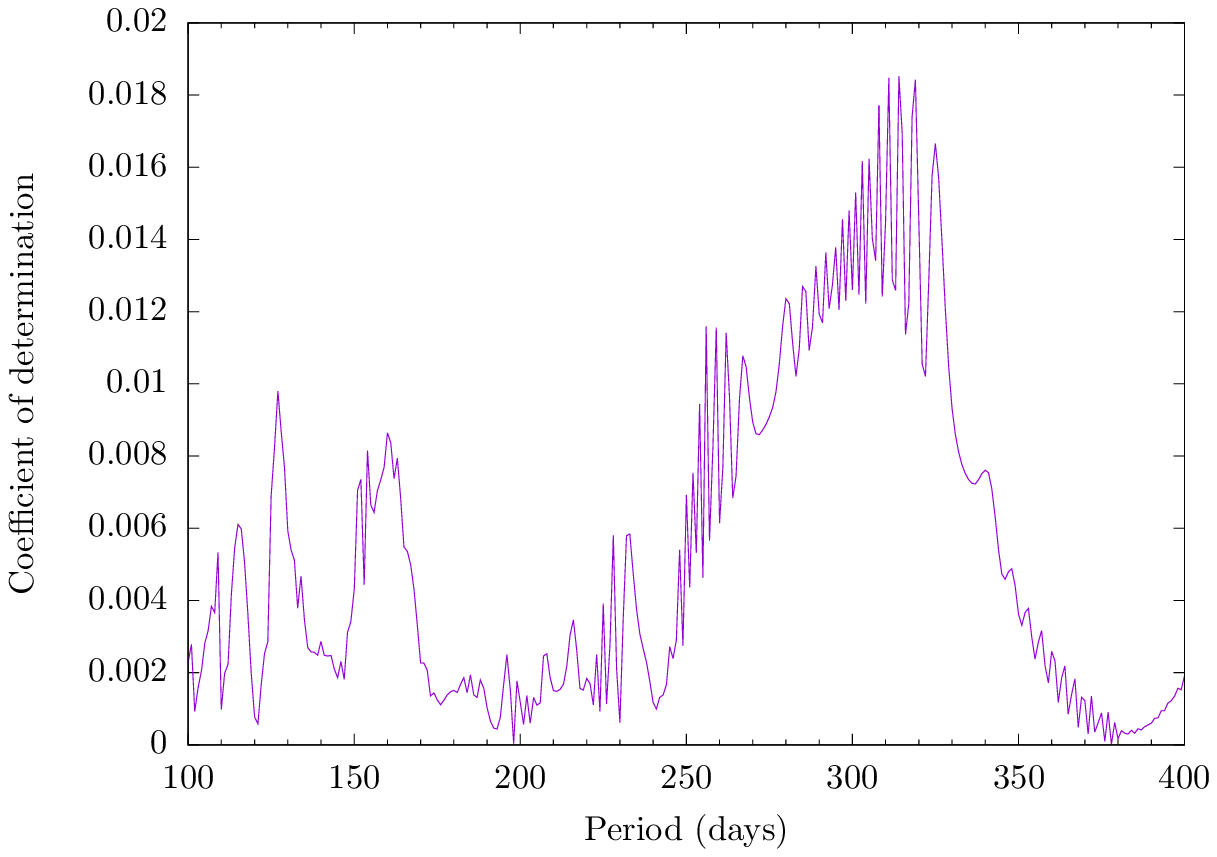}
  \caption[Figure]{From left to right, the figure shows the optical, X and 
  $\gamma$-ray lightcurves of the blazar Mrk~421 (top row).  The horizontal
  solid lines mark the mean flux levels and the dotted ones show the \(
  3-\sigma \) level above the mean, below which the outburstless state is
  defined.  Similarly, the
  middle row presents the corresponding spectral window functions \( W  \)
  \citep[see e.g.][]{dawson10} associated to the time samplings of the 
  databases.  The bottom row presents the periodograms obtained using RobPer 
  for each waveband.  The plots on the last two rows were 
  produced using the outburstless data.
  }
  \end{center}
\label{fig01}
\end{figure*}

  From the RobPer analysis  we have found a consistent periodicity in the
outburstless signal with period of order \( \sim 310\ \text{days} \).
In order to verify this result independently, we implemented a maximum
likelihood analysis~\citep[e.g.][]{abuelo} for a normalised Gaussian
noise observed through a window eclipse function \( e(t) \) for the
high quality hard X-ray data.  The lack of errors in the optical data,
plus the shorter temporal extent of the $\gamma$-ray  observations
make the X-ray periodogram the one where the 310 feature is clearest.
Therefore, the band we use for the further maximum likelihood analysis is
the X-ray one.  Whenever \( \sin( \omega t + \phi \) ), being \( \omega
\), \( t \) and \( \phi \) the frequency, time and phase respectively,
exceeds a critical threshold parameter \( \Theta \) such that \( -1 \leq
\Theta \leq 1 \), the signal is reduced by an eclipse fraction \( f \),
otherwise the inherent Gaussian noise is unaffected, i.e.:

\begin{equation}
  e\left(t\right)=\left[1-\frac{\left|\Theta-\sin\left(\omega
    t+\phi\right)\right| f }{\left|1-\Theta\right|}\right].
\label{eq01}
\end{equation}

  The log-likelihood function is now given by:

\begin{equation}
  \ln\mathcal{L}\left(T,\phi,f,\Theta;\left\{S_i, \Delta S_i \right\}\right) = \sum_i \frac{ \left( S_i / e(t)
    - 1 \right)^2  }{ 2 \Delta S_i^2 },
\end{equation}

\noindent where \( S_i \) and \( \Delta S_i \) are the normalised \(i\)-th
measured signal value and its uncertainty, respectively, with a period
\( T :=  2 \pi / \omega \).  A dense exploration of parameter space
returns the following optimal parameters: \( T = 310.54 \pm 1 \), \(
\phi = 1.52 \pm 0.02\), \( f =-0.0015 \pm 0.002 \) and \( \Theta =
0.954 \pm 0.002 \).  Given the X-ray data mean of \( 0.0011 \), the
percentage eclipse fraction is given by \( 136.4 \pm 20 \).

  The remarkable conclusion is that $f<0$. Therefore the eclipse in fact
corresponds to a slight brightness enhancement. It is tempting to identify
this with the gravitational lensing effect of an eclipse by a super
massive black hole companion, which would be consistent with the short
inferred fractional duration of the eclipse of $\left(\arcsin(0.954\pm
0.002) - \pi/2\right) / \pi = 9.7 \pm 0.2  \%$.  We see that a full
maximum likelihood parametric modelling of the light curve allows to
go beyond just periodicity, to gain a deeper insight into the physical
characteristics of the system studied.

  Note that even though a full sine curve was allowed by the maximum
likelihood analysis, the optimal eclipse does not resemble a sine curve.
This fact highlights the importance of considering very general periodic
functions in the original periodogram analysis, as is the case for the
RobPer package used here.

\section{Discussion}
\label{discussion}

  An established fact of Fourier periodogram analysis is that it cannot
deal with irregularly sampled time series or with highly variable measurement
accuracies~\citep{anita01}, such as the optical light curves used in
this work.  To avoid the use of a Fourier frequency based analysis on
the data, we used the R-package RobPer~\citep{anita02}.


The high frequency oscillation modes in all the light curves analysed in
this work follow a white-like noise pattern and are thus not related to
any self-organised phenomenon \citep{Landa}.  This behaviour corresponds
to  microphysical activity of the source, producing non-correlated
variability, typical of random stochastical processes.  When taking
into account both active (with outbursts) and outburstless states of
the light curves, the low frequency oscillation modes show a flicker
pink-like noise for the active state and a brown-like noise pattern in
the outburstless state.  The idea of analysing the outburstless signal in
the search for periodicities is inspired by the fact that Mrk~421 has a
self-organised flicker noise in the low frequency regime.  This is quite
probably associated to the emissions from the inner central engine,
contrary to what is found in the quasar PG~1302-102 which exhibits
a W-damped random walk behaviour \citep{Graham} associated with a
brown-like noise.  The \( 1/f \)  noise is the result of an infinite
superposition of stochastic signals \citep{Eliazar1,Eliazar2} and so,
it has no main periodicities associated.  In order not to confuse our
calculations with any false periodicity in this source, we removed
any signal above the \( 3 \)-\(\sigma\) noise level.  With this,
a brown-like noise in the resulting outburstless state is obtained.
Thus, the existence of a highly correlated process such as the inferred
eclipsing supermassive binary black hole system appears natural, as in
the case of the PG~1302-102 quasar mentioned above.

  We conclude with Figure~2 which shows the folded X-ray
light curve using a folding period of 310 days (approximately a  10th of
the temporal extent of the data) with corresponding stacked error bars,
together with the optimal generalised eclipse model.  Notice that having
close to 10 cycles allows to appreciate directly the brightening
identified by the maximum likelihood model, despite the extremely noisy
data.  Our results can be interpreted as an eclipse produced by a
supermassive black hole orbiting the central engine resulting in a
gravitational lensing brightness enhancement.

\begin{figure}
  \begin{center}
  \includegraphics[scale=0.70]{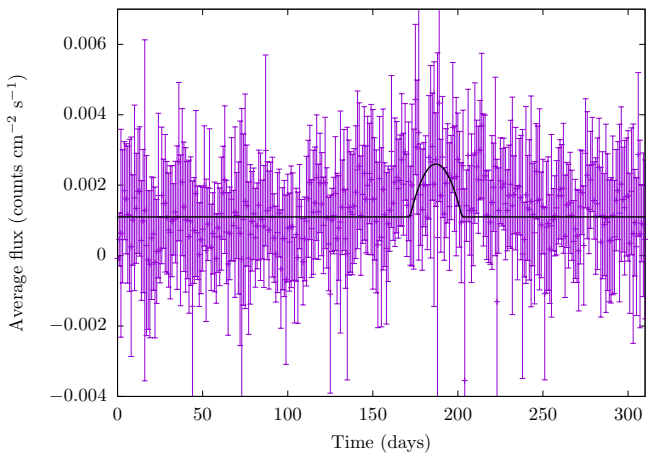}
  \caption[Figure]{ The figure shows the folded light curve of 
  the X-ray data using a period
  of 310 days, with average error bars. The solid curve is the best
  fit generalised eclipse found using the maximum likelihood model
  described, having a peak enhancement of \( 136 \%\).
  }
  \end{center}
\label{fig02}
\end{figure}

\section*{Acknowledgements}
We thank an anonymous referee for his valuable comments, which resulted
in an enhanced and more complete version of this letter.  We thank Anita
Monika Thieler for her invaluable help on better understanding the
R package RobPer.  This work was supported by DGAPA-UNAM grants
IN112616, IN111513, IN111514, IN100814, and by CONACyT grants
240512, 101958.  OL and FDL-C gratefully acknowledge a postdoctoral
DGAPA-UNAM grant.  OL thanks Capital Semilla program at IBERO.  
FDL-C gratefully acknowledges financial support
from Colciencias (Programa Es Tiempo de Volver) and Vicerrector\'{\i}a
de Investigaci\'on y Extensi\'on, Universidad Industrial de Santander,
grant number 1822. EB, JIC, NIF, XH, OL, FDL and SM acknowledge economic
support from CONACyT (13564,50102,53704,25006,62929,57585,26344). We thank
the AAVSO International Database contributed by observers worldwide and
used in this work for the optical light curve of Mrk~421. We also thank
the FERMI collaboration for the public database used in this work for
the \(\gamma\)-ray data.




\bibliographystyle{mnras}
\bibliography{mk} 

\bsp	
\label{lastpage}
\end{document}